\documentclass[sigconf]{acmart}

\usepackage{xspace}

\newcommand{\shifter}{\textit{Shifter}\xspace}
\newcommand{\ligo}{\texttt{LIGO}\xspace}
\newcommand{\atlas}{\texttt{ATLAS}\xspace}
\newcommand{\nanograv}{\texttt{NANOGrav}\xspace}
\newcommand{\powergrid}{\texttt{PowerGRID}\xspace}
\newcommand{\pyscf}{\texttt{PySCF}\xspace}
\newcommand{\qwalk}{\texttt{QWalk}\xspace}
\newcommand{\qespresso}{\texttt{QuantumEspresso}\xspace}
\newcommand{\insilico}{\textit{in silico}\xspace}
\newcommand{\via}{\textit{via}\xspace}

\newcommand{\etc}{\textit{etc.\xspace}}

\newcommand{\mpi}[1]{\texttt{MPI\_{#1}}\xspace}

\usepackage{booktabs}
\graphicspath{{figures/}{acmart-master/}}
\setcopyright{acmlicensed}

\acmDOI{10.1145/3219104.3219145}

\acmISBN{978-1-4503-6446-1/18/07}

\acmConference[PEARC '18]{Practice and Experience in Advanced Research Computing}{July 22--26, 2018}{Pittsburgh, PA, USA}
\acmYear{2018}
\copyrightyear{2018}

\acmPrice{15.00}

\acmBooktitle{PEARC '18: Practice and Experience in Advanced Research Computing, July 22--26, 2018, Pittsburgh, PA, USA}

\begin{document}
\title[Container solutions for HPC systems]{Container solutions for HPC Systems: A Case Study of Using Shifter on Blue Waters}

\author{Maxim Belkin}
\affiliation{
  \institution{University of Illinois at Urbana-Champaign}
  \department{National Center for Supercomputing Applications}
  \city{Urbana}
  \state{IL}
  \postcode{61801}
  \country{USA}
  }
\email{mbelkin@illinois.edu}
\author{Roland Haas}
\affiliation{
  \institution{University of Illinois at Urbana-Champaign}
  \department{National Center for Supercomputing Applications}
  \city{Urbana}
  \state{IL}
  \postcode{61801}
  \country{USA}
  }
\email{rhaas@illinois.edu}
\author{Galen Wesley Arnold}
\affiliation{
  \institution{University of Illinois at Urbana-Champaign}
  \department{National Center for Supercomputing Applications}
  \city{Urbana}
  \state{IL}
  \postcode{61801}
  \country{USA}
  }
\email{gwarnold@illinois.edu}
\author{Hon~Wai Leong}
\affiliation{
  \department{National Center for Supercomputing Applications}
  \institution{University of Illinois at Urbana-Champaign}
  \streetaddress{1205 W. Clark St}
  \city{Urbana}
  \state{IL}
  \postcode{61801}
  \country{USA}
  }
\email{hwleong@illinois.edu}
\author{Eliu~A. Huerta}
\affiliation{
  \department{National Center for Supercomputing Applications}
  \department{Department of Astronomy}
  \institution{University of Illinois at Urbana-Champaign}
  \streetaddress{1205 W. Clark St}
  \city{Urbana}
  \state{IL}
  \postcode{61801}
  \country{USA}
  }
\email{elihu@illinois.edu}
\author{David Lesny}
\affiliation{
  \department{Department of Physics}
  \institution{University of Illinois at Urbana-Champaign}
  \streetaddress{1110 W. Green St}
  \city{Urbana}
  \state{IL}
  \postcode{61801}
  \country{USA}
  }
\email{ddl@illinois.edu}
\author{Mark Neubauer}
\affiliation{
  \department{Department of Physics}
  \institution{University of Illinois at Urbana-Champaign}
  \streetaddress{1110 W. Green St}
  \city{Urbana}
  \state{IL}
  \postcode{61801}
  \country{USA}
  }
\email{msn@illinois.edu}


\renewcommand{\shortauthors}{M. Belkin et al.}

\begin{abstract}
Software container solutions have revolutionized application development
approaches by enabling lightweight platform abstractions within the so-called
``containers.'' Several solutions are being actively developed in attempts to
bring the benefits of containers to high-performance computing systems
with their stringent security demands on the one hand and fundamental resource
sharing requirements on the other.

In this paper, we discuss the benefits and short-comings of such solutions when
deployed on real HPC systems and applied to production scientific
applications. We highlight use cases that are either enabled by or
significantly benefit from such solutions. We discuss the efforts by HPC
system
administrators and support staff to support users of these type of workloads on
HPC systems not initially designed with these workloads in mind focusing on
NCSA's Blue Waters system.
\end{abstract}



\maketitle

\section*{The rise of Containers}
\label{intro}

The enormous growth of computing resources has forever changed the landscape
and pathways of modern science by equipping researchers with the apparatus that
is impossible to realize experimentally. The great examples are data- and
compute-enabled machine and deep learning algorithms that control self-driving
cars; precise \insilico studies of complete virus capsids that further our
understanding of their pathogenic pathways; and the fascinating studies of
gravitational waves that resonate around our Universe.

This growth of computing resources has been multi-directional: they increased
in their availability, performance, and assortment.  A typical computer today
is equipped with Graphics Processing Units (sometimes combined with Central
Processing Units), abundant Random Access Memory, hard drives that can store
Tera bytes of data, and many other, often highly specialized, hardware.
Supercomputers, the high-performance computing (HPC) resources that drive
modern science, have additional levels of complication with their stringent
security demands, fundamental resource sharing requirements, and many
specialized libraries that enable use of the underlying hardware at its peak
performance.

Variations in hardware and software stacks across leadership-class computing
facilities have raised a great deal of concern among researchers with the most
prominent one being \textit{reproducibility} of computational studies. To
ensure reproducibility, it is critical to use portable software stacks that can
be seamlessly deployed on different computing facilities with their specific
architectures. This need has driven the development of software solutions that
abstract the underlying hardware away from the software.  Today's most popular
examples include container solutions like Docker, virtual machine solutions
such as VirtualBox and VMWare Workstation, and others.
These solutions differ in levels at which abstractions take place (hardware,
OS, \etc), abstracted and required resources, as well as all auxiliary
tools that together comprise their \textit{ecosystems}. In this article we
focus on a container solution for HPC systems: \shifter.

In addition to facilitating the use of complex software stacks within the HPC
community, containers have also played a central role in a new wave of
innovation that has fused HPC with high-throughput computing (HTC)---a
computing environment that delivers a large amount of computing power over
extended periods of time. A number of large scientific collaborations have
made use of containers to run computationally demanding HTC-type workflows using
HPC resources.

In this paper, we showcase a number of efforts that have successfully harnessed
the unique computing capabilities of the Blue Waters supercomputer, the
NSF-supported, leadership-class supercomputer operated by National Center for
Supercomputing Applications (NCSA), to enable scientific discovery. We focus on
efforts that have been spearheaded by researchers at NCSA and the University of
Illinois at Urbana-Champaign~\cite{BOSS:2017,2016CQGra..33u5004U}. These
efforts provide just a glimpse of the wide spectrum of applications in which
containers help advance fundamental science: from the discovery of
gravitational waves from the collision of two neutron stars with the \ligo and
Virgo detectors~\cite{bnsdet:2017}, to the study of the fundamental building
blocks of nature and their interactions with CERN's Large Hadron Collider
(LHC).


\section*{Shifter on Blue Waters}
\label{shifteronbw}

\begin{figure}[htbp]
\centering
\includegraphics[width=\columnwidth]{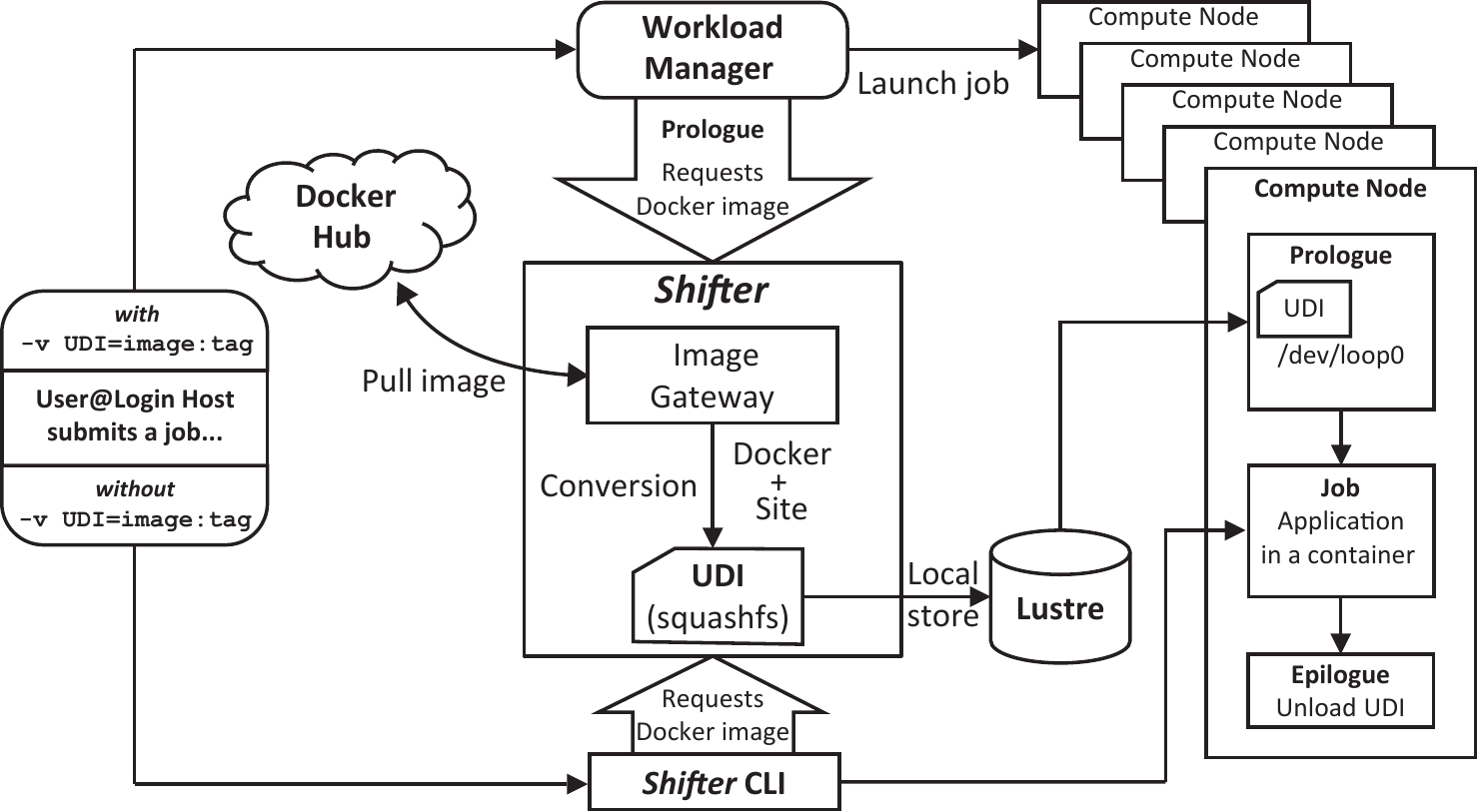}
\caption{Architecture of Shifter implementation on Blue Waters}
\label{shifter-flow}
\end{figure}

\shifter~\cite{shifter:cug2016} is a container solution that is designed
specifically for HPC systems and which enables hardware abstraction at the
OS level.
Figure~\ref{shifter-flow} illustrates the workflow of a typical
\shifter \textit{v}16.08.3 job on the Blue Waters supercomputer.

It is extremely easy to get started using \shifter as all one has to do is
provide an additional generic resource request: \texttt{-l gres=shifter16},
either on the command line or in a job batch script.
This PBS directive is mandatory for any \shifter job on Blue Waters
as it instructs system's workload manager to
execute special \textit{prologue} and \textit{epilogue} scripts before and after
job execution in order to set up and tear down the container environment on all
compute nodes.

The other immediate advantage of \shifter is that it can work with Docker
images---one of the most popular container formats---out of the box.  There are
two ways to specify which container to use in a batch job: either as a PBS
directive \texttt{-v UDI=<image:tag>} or as an argument \texttt{--image=<image:tag>}to
the \texttt{shifter} command provided by \shifter.  In both cases, \texttt{image} corresponds
to \texttt{repository/imagename} on Docker Hub. 

When UDI is specified as a PBS
directive, the prologue script communicates with the \shifter image gateway to
check that the image exists and download it if it does not. The image
gateway then applies site-specific environment changes to the image
and converts it into a \textit{squashfs}-formatted image file, typically
referred to as User Defined Image, or UDI. The prologue script then proceeds to
mount the UDI on all compute nodes allocated to the job. The UDI image
file is stored on the Lustre file system and subsequent jobs 
requesting the same image can use it without repeating all of the above
steps.
Upon completion of such as job, the epilogue script unmounts the UDI from the compute nodes
and performs site-specific procedures necessary to properly clean up
the environment on compute nodes.
Both, site-specific environment changes to the downloaded image and cleanup
procedures are specified by the system administrators.

The alternative way to specify UDI is by supplying it as an argument to the
\texttt{--image} flag of the \texttt{shifter} command.
In combination with Blue Waters' Application Level Placement Scheduler (ALPS)
task launcher,
to run an application within a container environment one can use the following
command: \begin{verbatim} $ aprun -b -- shifter --image=<image:tag> --
<application> \end{verbatim}
The \texttt{shifter} command above initiates a series of operations which are similar
to those executed by the \textit{prologue} script of the workload manager.
However, it not only provides the option to select container environment
``on-the-fly,'' but also allows \shifter users to use several different images
within the same job! 
This method
is arguably best suited for single-task applications. For containerized MPI
applications it is still recommended to use PBS directives to set up the
container environment on all compute nodes.

The core image gateway manager of \shifter is designed as a RESTful service.
It is written in Python language and depends on multiple software components:

\textit{Flask} - A Python-based framework that provides a RESTful API as an
interfacing layer between user requests and the underlying image gateway. The
use of RESTful API replaces local Docker engine as a gateway for users to
request containers from a Docker registry.

\textit{MongoDB} - A distributed database to store metadata of available
container images and their operational status: whether the image is still being
downloaded, its conversion status, its readiness to use, or any failure
encountered.

\textit{Celery} - A Python-based asynchronous and distributed task queueing
system to service user requests. \textit{Celery} provides better scalability for
multiple requests through queueing and dispatch to a distributed pool of
workers.

\textit{Munge} - An authentication service for creating and validating
credentials, designed to be highly-scalable, which is ideal for high-performance
computing environment. \shifter uses \textit{Munge} to authenticate user
requests from clients to the gateway manager.

\textit{Redis} - An in-memory data structure store, used as a database, cache
and message broker to support \textit{Celery}'s functionality. It captures the
operational state of the \shifter image gateway service to allow live
reconfiguration of \shifter (service restart) without interrupting any current
operations.

When compared to its previous version~\cite{shifter:cug2015}, \shifter
\textit{v}16.08.3 features improved functionality and
performance. Yet, just like the predecessor, it still introduces a noticeable overhead for system
administrators who are responsible for its back-end operation because it relies
on a number of very different components working seamlessly and with no
interruptions. Without a doubt, it is much harder to troubleshoot an issue that
involves \shifter as its root cause may not come from the tool itself but from
one of its dependencies. 

During the production use of \shifter on Blue Waters,
the following issues have been encountered:

1. \textit{Stale ``PENDING'' state}.
When downloading containers from Docker registry, the status would stay in
``PENDING'' state indefinitely until its metadata is manually deleted from
MongoDB's database. This usually happens when a user aborts the download of
a large container from the Docker registry before the download completes.

2. \textit{False ``READY'' state}.
Status of a container image would indicate ``READY'', even though \shifter has, in
fact, failed to mount the UDI on the compute nodes due to the unfinished download of
the employed container image. The troublesome UDI file has to be removed from the storage
and Docker image has to be re-downloaded.

3. \textit{Persistent out-of-memory issue on gateway host}.
There was an incident when the gateway manager caused the gateway host to run
out of memory and, consequently, go down because multiple threads were
downloading the same image from the Docker registry. Upon rebooting the host and
restarting all of the services required by \shifter, multiple threads resumed
their downloads leading to repeated failures. Subsequent restarts did not
produce expected results. The solution was to remove the Redis dump DB file.

4. \textit{Failure to mount UDI when Munge is not running}.
Munge is crucial for \shifter to function properly.  A compute node that does
not have Munge service running would not be able to authenticate with the
\shifter image gateway and thus would fail to mount a UDI.

5. \textit{Failing to run at scale}.
The major challenge that we had to address on Blue Waters was to make
\shifter jobs run at scale. The issue was caused by the
bottleneck in \texttt{getgrouplist} and
\texttt{getgrgid} functions that \shifter uses to set up the containers on
compute nodes. These two functions query local passwd and group files and LDAP.
Because Blue Waters does not store regular user and group information in
passwd and group files, \shifter was trying to get the
\texttt{gid}s of the executing user from LDAP. For jobs with a large node count this step results in a
large number of concurrent requests being sent to the underlying LDAP server. As a
result, not all requests receive a response from the server. To work
around this issue, we had to turn on the Name Service Cache Daemon (NSCD) service 
on all compute nodes allocated to \shifter jobs. The NSCD service caches LDAP
entries on the compute nodes and, therefore, enables their fast lookup.


\section*{MPI applications in \shifter jobs}
\label{mpi}

MPI is a performance-critical component of and \textit{de facto} the standard
for writing applications that run at scale. Therefore, for systems like Blue
Waters it is crucial to understand the overhead that applications  within
\shifter UDIs have to pay in order to run on multiple nodes. To estimate this
overhead, we compared \shifter to the Cray Linux Environment (CLE) using the
OSU Micro-Benchmarks.

\begin{figure}[htbp]
  \centering
  \includegraphics[width=\columnwidth]{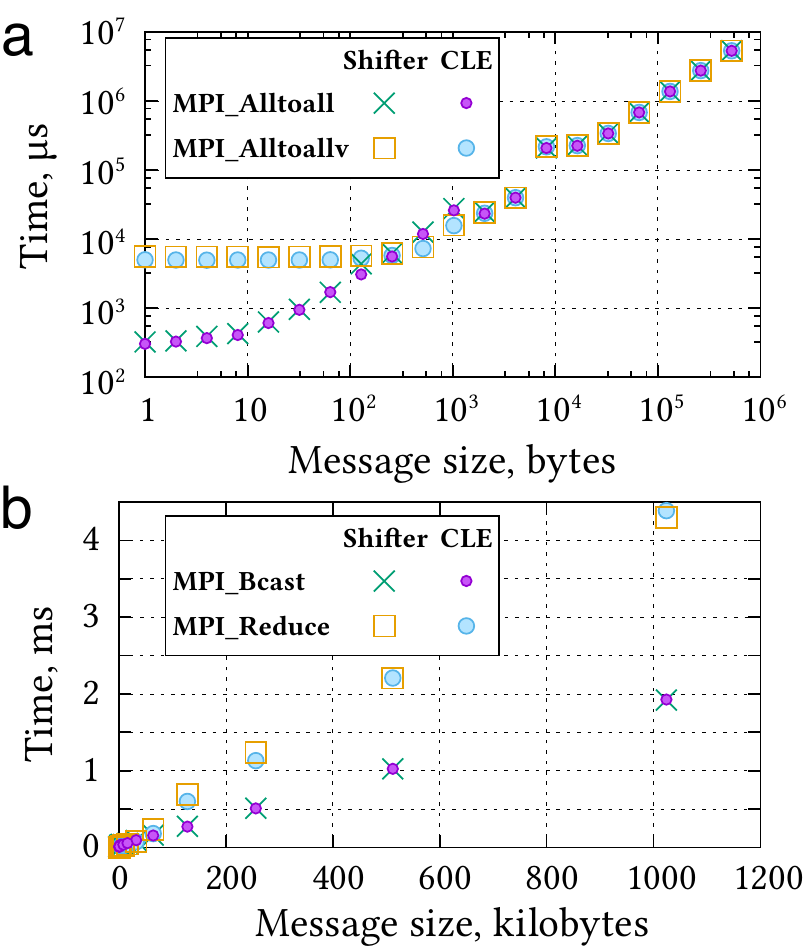}
  \caption{Comparison of \textbf{OSU micro benchmarks'} results measuring
  MPI performance in \shifter and Cray's native Linux Environment
  on Blue Waters using 64 compute nodes and 1,024 MPI Ranks.
\textbf{(a)} \texttt{MPI\_Alltoall} and \texttt{MPI\_Alltoallv}.
\textbf{(b)} \texttt{MPI\_Bcast} and \texttt{MPI\_Reduce}.
}
\label{64nodes}
\end{figure}

A selection of representative benchmarks were run: \mpi{Bcast}, \mpi{Reduce},
\mpi{AlltoAll}, and \mpi{AlltoAllv}. Tests were performed on 64 and 1024 ranks,
that correspond to 4 and 64 compute nodes on Blue Waters, correspondingly, see
Figure~\ref{64nodes}.  Employed \shifter image was based on clean Centos 7
Docker image, with MPICH~\textit{v}3.2 and
OSU Micro-Benchmarks~\textit{v}5.3.2 installed from source. Our results
suggest that MPI
performance in CLE and \shifter is statistically the same. This stunning
result is not surprising, however, because \shifter is able to use the Cray MPI
low level communication libraries through the MPICH ABI compatibility
initiative.

We set up MPI benchmarks in a way that made \shifter the only component that
could significantly affect the results. In particular, the binaries were built
with tools provided by the GNU Programming environment on Blue Waters
(\textit{PrgEnv-gnu}) for the CLE tests, and with \texttt{mpicc}  that calls
GNU compilers in the \shifter UDI that was based on Centos7 Docker image.
Tests were run from the same batch job, minimizing the effect of
node placement and Gemini network paths on the obtained results as much as possible.
Only the
variable network traffic that is associated with the production machine and
that we don't have control over could have impacted the results.  Because the
results were obtained from the same jobs, we're confident that they are
valid and reproducible.


\section*{I/O Performance in \shifter jobs}
\label{io}
\begin{figure}[htbp]
\centering
\includegraphics[width=\columnwidth]{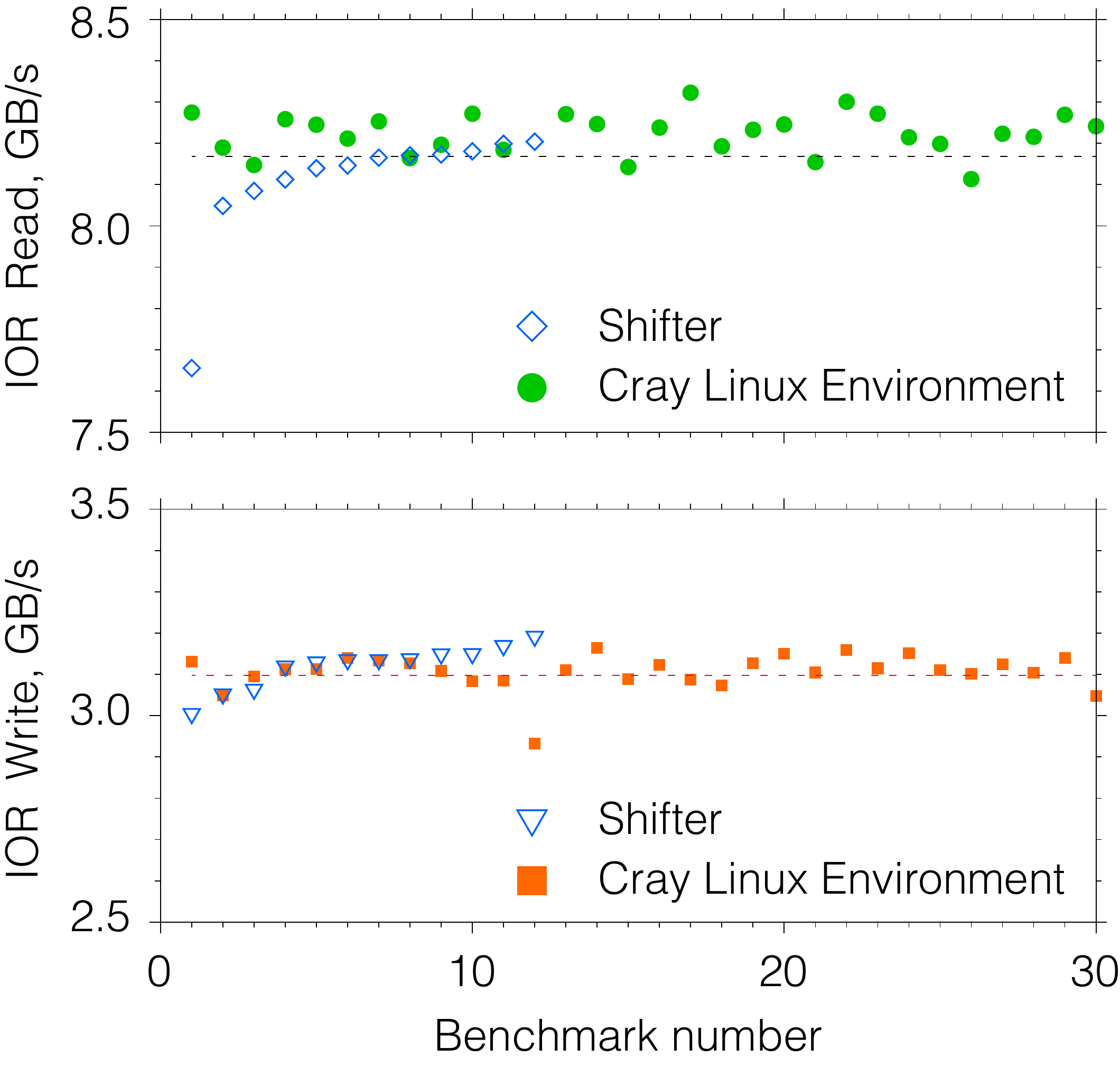}
\caption{Comparison of \textbf{IOR benchmark} results of
  IO performance in \shifter and Cray's native Linux Environment
    on Blue Waters using 16 compute nodes and 7 cores for reading and
    writing operations on each node.}
\label{ior}
\end{figure}

Performance of read and write operations is crucial for the HTC type of
applications that deal with lots of data. To see if \shifter imposes any
input/output (I/O) overhead, we ran the IOR MPI I/O benchmark
(https://github.com/hpc/ior,  commit \texttt{aa604c1}) using 16 nodes and 7
cores for reading and writing operations on each node.
Blue Waters runs the IOR benchmark on a regular schedule using the
Jenkins testing infrastructure. To make the comparison between the tests
meaningful, we used the same input and node layout in our \shifter tests.
Our results suggest, that there is no substantial differences between I/O
performance in the native Cray Linux Environment (Jenkins test case) and the
\shifter case, see Figure~\ref{ior}.

\section*{Start-up time of \shifter jobs}
\label{startup}

\begin{figure}[htbp]
\centering
\includegraphics[width=1.0\linewidth]{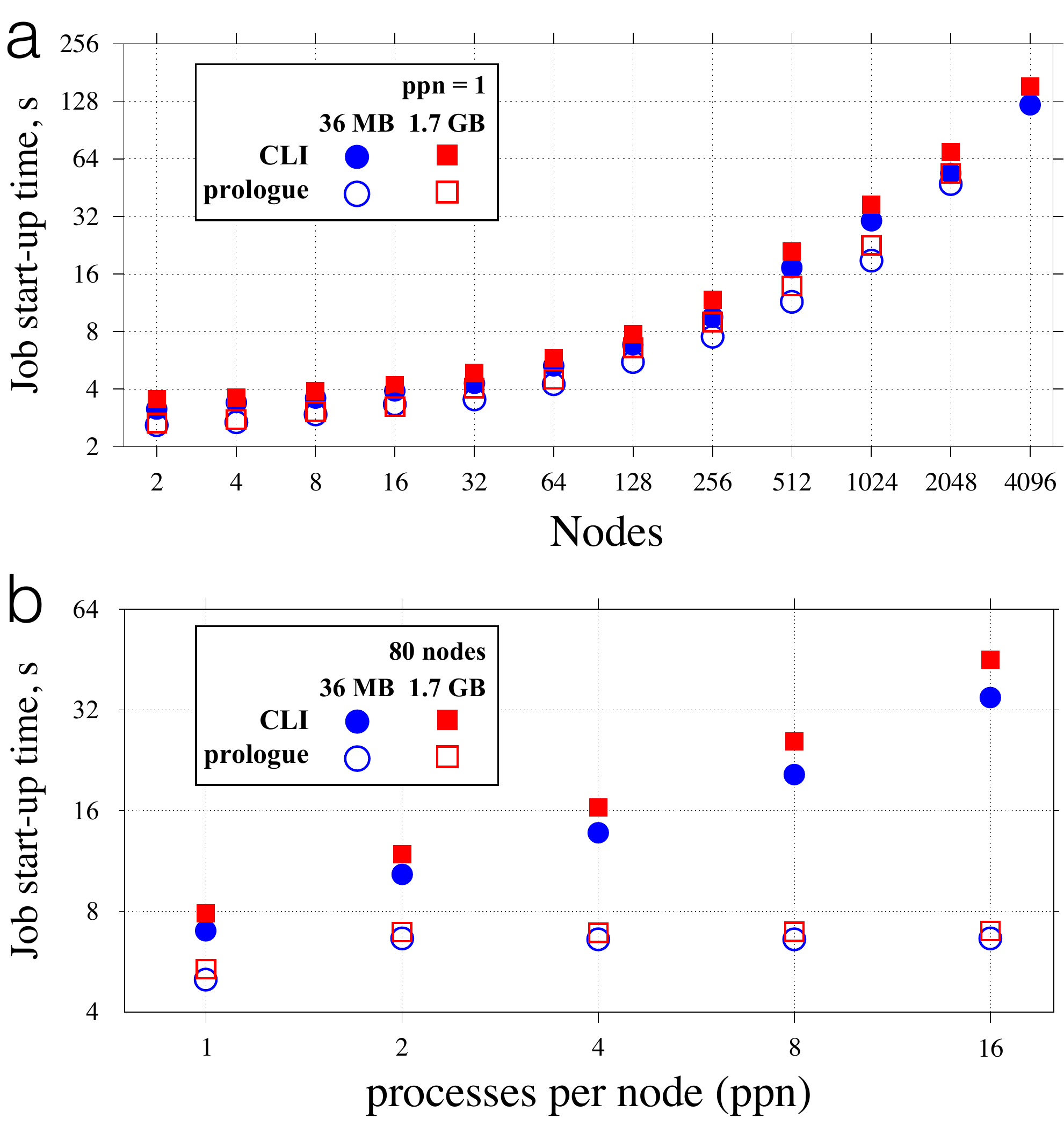}
\caption{Start-up time of \shifter jobs on Blue Waters.  (a) Dependence of a
  \shifter job start-up time on the number of nodes. Start-up time is found to
  be practically independent of the way we specify which UDI to use in a job and
  the size of that image.  (b) Dependence of the start-up time of a \shifter job
  that uses 80 nodes on the number of MPI processes used on each node. When UDI
  is specified at the time the job is submitted, job start-up time does not
  change when with the number of MPI processes used on each node!
  }
\label{fig:startup-time}
\end{figure}

\shifter enables many new types of applications take advantage of HPC resources.
As such, one may expect untraditional for HPC usage patterns to emerge. For
example, starting production simulations or different stages of analysis from
within a \shifter image multiple times within a job. To help users with such
applications better utilize HPC resources, we analyzed the start-up time of
\shifter jobs for User-Defined Images of two sizes: 36~MB and 1.7~GB. The
results are shown in Figure~\ref{fig:startup-time}. 

We investigated how start-up time of a \shifter job depends on the number
of nodes used by the job that exploits only 1 processor on each node. In our
tests, we started \shifter jobs in two different ways: 1. by specifying UDI at
the time the job was submitted, and 2.  by specifying UDI as an argument to the
\texttt{shifter} command. All of our tests suggest that start-up time of a
\shifter job is practically independent of the size of the User-Defined Image!
However, we find that for jobs using less than 256 nodes, the dependence of the
start-up time on node count is \textbf{sublinear}, beyond 256 nodes the
dependence becomes linear, and beyond 2,048 -- superlinear,
see~Figure~\ref{fig:startup-time}\,a.

We also studied the dependence of \shifter job start-up time on the number of
MPI processes used on each node. All of these tests were performed using 80
compute nodes. And again, we find start-up time to be practically independent of
the size of the User-Defined Image we use. However, we find that when we specify
UDI at the time we submitted the job, \texttt{aprun} calls take the same amount
of time regardless of the number of processes on each node we request. This
behavior is opposite of what we observe when we specify UDI as an argument to the
\texttt{shifter} command. This observation suggests that if multiple calls to
applications within the same UDI are necessary in a single job, it is advisable
to specify UDI at the time the job is submitted.

\section*{Codes using \shifter on Blue Waters}
\label{shifter_users}

\shifter was added to Blue Waters system in September of 2016 and was
first used in a production simulation in January of 2017 by the ATLAS project
to analyze data from the CERN's Large Hadron Collider~\cite{bafz-project}.
The science team worked with the Blue Waters project to set up and test \shifter.
The tested version of \shifter was then officially presented in a monthly user call in
February of 2017~\cite{shifter-presentation:2017}.

\begin{figure}[htbp]
\centering
\includegraphics{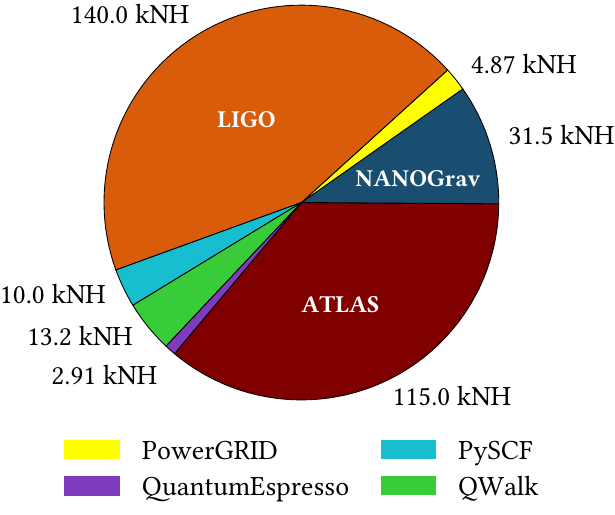}
\caption{Thousands of node-hours consumed by different codes using \shifter in
  the period 2016/09 -- 2017/03. Of the codes shown, \atlas, \nanograv, and
  \ligo are well established high throughput computing workflows. \pyscf, \qwalk
  and \qespresso are traditional HPC codes that employ MPI to achieve
  parallelization.  \powergrid is a GPU-enabled MPI code that
  can employ multiple GPUs on the compute nodes.
  }
  \label{fig:hour_pie_graph}
\end{figure}

\begin{figure}[htbp]
\centering
\includegraphics{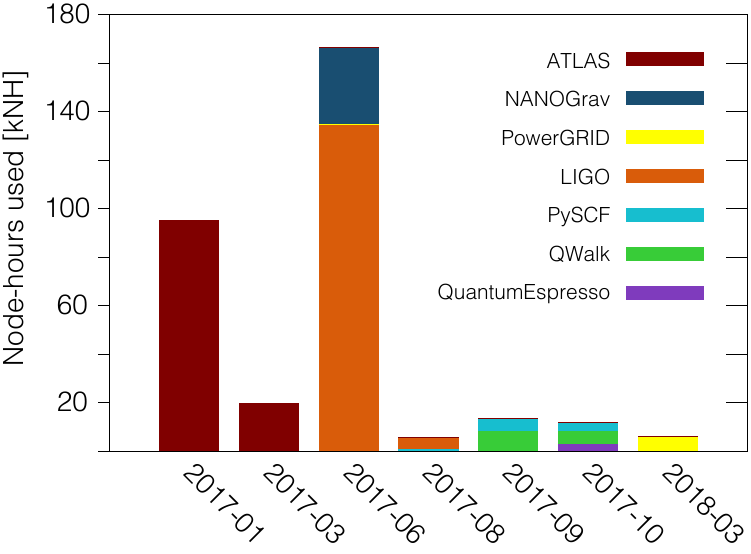}
\caption{
  Node-hours used by \shifter-enabled codes on Blue Waters since 2016.
  The three early adopters---\atlas, \nanograv, and
  \ligo---employ typical for HTC workflows with multiple pilot jobs and
  \texttt{HTCondor} serving as the main workflow mananger.
  \pyscf, \qwalk and \qespresso are
  traditional for HPC computational physics and chemistry codes.
  \powergrid is a new MPI- and GPU-enabled code for MRI image reconstruction.
  }
\label{fig:hour_graph}
\end{figure}

In order to learn about the codes that benefit from \shifter on Blue Waters, we
collected information about its usage by analysing accounting records for the
period from September, 2016 to March, 2017. In our analysis we did not include
the simulations that ran for less than 1 hour. Interestingly enough, however,
we found no significant difference in the distribution of codes when using a 5
minute ``cutoff'' instead.  Figure~\ref{fig:hour_pie_graph} shows the
distribution of node-hours consumed by different codes during the analyzed
period.
As is clear from
Figure~\ref{fig:hour_pie_graph} the majority of the node-hours used with \shifter
were consumed by \atlas, \nanograv, and \ligo projects.
All of them are HTC codes that employ a large number of short and
independent tasks that represent a trivially parallelizable workload.
On HPC systems like Blue Waters, such codes typically use the
so-called ``pilot jobs''~\cite{Luckow:2012:TCM:2287076.2287094} that reserve
compute nodes and aggregate them to a large shared compute pool of the
HTC workflow manager.
All three codes employ \texttt{HTCondor}~\cite{condor-practice} as the workflow
manager and scale well to a large number of nodes.
This scalability is achieved by using multiple pilot jobs to allow the workflow
manager to release compute nodes when there are not enough tasks to utilize
all provided resources.
\pyscf, \qwalk, and \qespresso represent
``traditional'' MPI-based HPC codes that utilize all
allocated compute nodes and, therefore, benefit from the \shifter's ability to
support MPI from within the containers.
Finally, \powergrid~\cite{cerjanic:16:pao}  is a modern, multi-GPU MPI code
for reconstructing images obtained with the Magnetic Resonance Imaging technique.
Figure~\ref{fig:hour_graph} shows the distribution of node-hours used each
month among the codes.
As can be seen from Figure~\ref{fig:hour_graph}, \shifter has \textit{not} been
used continuously by any single code or science group on Blue Waters. For codes
such as \nanograv and \ligo, this is due to the nature of their
discrete analysis ``Campaigns'' during which collected data is analysed.  Codes
such as \powergrid are still in the early stages of exploring the
capabilities of \shifter. A follow-up study is necessary to determine if the
observed non-continuous usage pattern is typical for applications that use \shifter.

Finally, Table~\ref{tab:nodes-per-code} shows the number of nodes used by
different applications on Blue Waters.
\begin{table}
\caption{
  Top science applications and projects that use \shifter on Blue Waters.
  Columns show the number of nodes used in a typical job (\textbf{Nodes}),
  number of jobs ran (\textbf{Frequency}), and the total charge for the jobs
  (\textbf{Node-Hours}). The top three science projects that consumed the most
  resources while using \shifter are \ligo, \atlas, and \nanograv.
}
\label{tab:nodes-per-code}
\begin{center}
\begin{tabular}{lrrr}
\toprule
  \textbf{Code} & \multicolumn{1}{l}{\textbf{Nodes}}
                & \multicolumn{1}{l}{\textbf{Frequency}}
                & \multicolumn{1}{l}{\textbf{Node-Hours}} \\
\midrule
\ligo       & $10$      & $8$       & $1,990$ \\
\ligo       & $50$      & $5$       & $2,650$ \\
\ligo       & $100$     & $1$       & $1,070$ \\
\ligo       & $500$     & $1$       & $6,030$ \\
\ligo       & $5,000$   & $4$       & $127,000$ \\
\atlas      & $16$      & $311$     & $115,000$ \\
\nanograv   & $1$       & $1,485$   & $28,500$ \\
\nanograv   & $100$     & $2$       & $3,010$ \\
\powergrid  & $800$     & $1$       & $4,870$ \\
\pyscf      & $1$       & $419$     & $10,000$ \\
\qwalk      & $1$       & $1,138$   & $13,200$ \\
\qespresso & $2$ & $422$ & $2,910$ \\
\bottomrule
\end{tabular}
\end{center}
\end{table}
As one can see from Table~\ref{tab:nodes-per-code}, most \shifter jobs are
small (16 nodes or less) with only \ligo and \powergrid
attempting to scale up to larger node counts.
This can be understood considering that available HTC tasks may not be
sufficient to keep thousands of cores busy. Yet, an HPC system can not
release just a fraction of nodes that are part of a job. This is the main
reason for using multiple pilot jobs that can be terminated when necessary.
The optimal size and number of pilot jobs depends on multiple factors such as
the latency of the HPC scheduler, the length of each task, the backlog
of available tasks in the workflow manager, and the ``cost'' of having idle
nodes.  Therefore, exploratory HTC runs use many small pilot jobs to
determine the optimum quantities while only a few large pilot jobs are then used for
production simulations, analysis, and testing.

\subsection*{\atlas, \nanograv, and \ligo}
\label{sec:htc-codes}

A lion's share of node-hours consumed by \shifter jobs on Blue Waters is
associated with the three big state-of-the-art research projects: \atlas,
\nanograv, and \ligo. Availability of sufficient computing resources was
crucial for their Nobel prize-winning works that detected the Higgs boson 
and gravitational waves in 2013 and 2017, respectively.

\begin{figure}[htbp]
\centering
\includegraphics[width=\columnwidth]{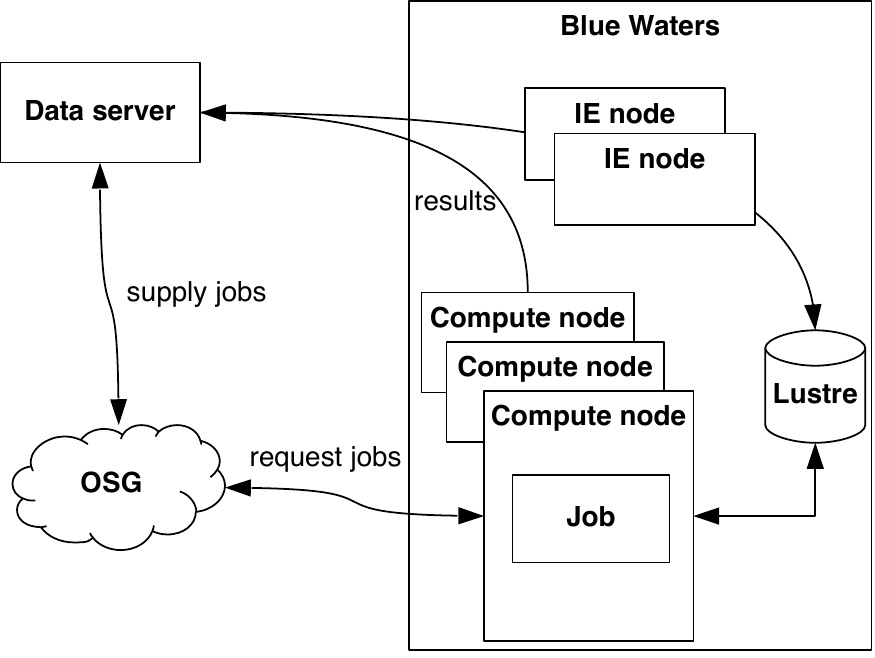}
  \caption{Interaction between an science project data repository, the Open
  Science Grid and Blue Waters. Import/Export (IE) nodes are Blue Waters'
  dedicated nodes that are used for file transfer.
  Figure reproduced from \cite{BOSS:2017}.}
  \label{fig:osg_interaction}
\end{figure}

Because all three codes use an Open Science Grid
(\texttt{OSG})~\cite{pordes2007open}-derived workflow, the challenges they face
and behaviour they exhibit are very similar.  Figure~\ref{fig:osg_interaction}
shows a typical setup when using Blue Waters and \shifter as a compute resource
in the \texttt{OSG}.
For simplicity, we use \ligo as a stand-in for all three codes but
the setup is, essentially, identical for all three projects.

The LIGO Scientific Collaboration employs \texttt{HTCondor} to analyse the
data recorded by the LIGO detector, requiring that data from a repository in
Nebraska, USA is transferred to a computing facility for processing.
\shifter enabled \ligo collaboration to use an \texttt{OSG}-ready Docker image
on Blue Waters, eliminating the need to adapt the image for each resource
provider. This allowed \ligo to use an operating
system environment which is certified by the collaboration for a detection
campaign and that matches the environment found on \texttt{OSG} resource
providers: \texttt{CentOS} instead of Blue Waters' native \texttt{CLE}.

To register Blue Waters with the \texttt{HTCondor} scheduler as an \texttt{OSG}
site, pilot jobs had to use a modified version of the
GlideinWMS~\cite{sfiligoi2009pilot} tool.  In the \shifter UDI based on the
\texttt{OSG} Docker image, the tool was immersed in an \texttt{OSG}-like
environment and, therefore, could download and execute the \ligo analysis code
as usual.

A complication arose due to \texttt{OSG} using \texttt{CVMFS}~\cite{cvmfs} to
distribute application codes like \ligo to the resource providers. Because
\texttt{CVMFS} relies on \texttt{FUSE}~\cite{libfuse} and the latter is not
supported by the OS kernel on Blue Waters, a copy of the relevant sections of
\texttt{CVMFS}'s data hierarchy had to be stored on the Blue Waters'
\texttt{Lustre} file system which is accessible from within the \shifter job.

Finally, analysis task required a data file of approximately $400\,\text{MB}$ in
size which was downloaded using \texttt{GridFTP} and \texttt{XRootD} transport
protocols~\cite{2017Weitzel}. With \texttt{GridFTP} extra care was necessary not
to overwhelm the data server because each \texttt{GridFTP} connection requires a
heavy-weight runtime environment to be initialized on the data server.
\texttt{XRootD} on the other hand is designed specifically for \texttt{OSG}
workflows and handles multiple transfers more gracefully.

Using this setup, Blue Waters contributed approximately $8,000$ node-hours to
LIGO's second observation campaign, temporarily becoming the peak resource
provider, and approximately $50,000$ node-hours to the \atlas project in
2017~\cite{neubauer:2017}.

Future HTC codes that rely on \texttt{OSG} resources will definitely
benefit from the experiences gained and the groundwork laid by \atlas,
\nanograv, and \ligo on Blue Waters. With the help of \shifter, only minimal
modifications are required to enable such codes take advantage of Blue Waters,
providing a new pool of compute resources otherwise unavailable to HTC codes.

\subsection*{\qwalk, \pyscf and \qespresso}
\label{sec:compchem-codes}

\qwalk~\cite{qwalk} and \pyscf~\cite{pyscf} are an electron
structure and computational physics / chemistry codes. Since \qwalk uses
a Quantum Monte Carlo (QMC) method, it parallelizes trivially to refine its
predictions using additional instances of the simulation. As such, no complex
workflow manager was required and researchers were able to develop
automation framework for use with \shifter independently.

\subsection*{\powergrid}
\label{sec:powergrid}

\powergrid~\cite{cerjanic:16:pao} is an MPI applications for medical
magnetic resonance image reconstruction that can take advantage of
GPUs. It relies on MPICH ABI compatibility to use a
single executable compiled and dynamically linked with MPICH that runs
under Cray's MPI stack on Blue Waters. \powergrid employs
parallelization to process multiple snapshots in parallel using MPI to
farm out tasks to the cores available to the job. The per-rank code is
parallelized \via \texttt{OpenACC} targeting Blue Waters' NVIDIA Kepler GPUs.
\shifter enabled the team to build a complex software stack with multiple
compiler dependencies and \texttt{CUDA} support that they can deploy on a
variety of underlying hardware.


\subsection*{Outlook}
\label{sec:outlook}

For all applications discussed in this paper, \shifter played a critical role in
making their execution on Blue Waters \textit{possible}. But why do we not see
more examples like this? If we look closer at scientific applications in
general, we find little consistency in the way these applications are developed.
This lack of consistency leads to the use of an array of tools and packages that
make the process of building applications even in a controlled environment
provided by Docker very difficult. Even more so, building applications in a way
that would allow them to take full advantage of the hardware provided by
leadership-class computing facilities while maintaining container portability.
Thus, despite all the benefits that \shifter brings to the world of
High-Performance \& Throughput Computing, there is still room for improvement.

\section*{Conclusions}
\label{end}

We described the lessons learned and experiences gained while adopting
\shifter as a container solution on the Blue Waters supercomputer. We presented
a thorough and up-to-date report on its performance, functionality, issues encountered,
and also the benefits and new possibilities that it enables. While some challenges
remain to be solved (unsupported or chip-specific and incompatible
instructions), \shifter has already provided a long-awaited solution
that enabled the HPC community to run complex and atypical (for HPC) software stacks.
Essentially, \shifter enabled HPC centers like Blue Waters to imitate Cloud
infrastructure which is sought after by the HTC community. Over the
last year, Blue Waters users have been steadily ramping up the utilization of
\shifter.  In addition to providing seamless access to the unique computing
capabilities of Blue Waters to run HTC-tailored workflows, \shifter
has provided the means to further a wave of innovation that has fused
HPC and HTC resources to address grand computational
challenges across science domains. We have showcased recent applications of
\shifter that
demonstrate the new role containers are starting to play in
maximizing the versatility and flexibility of HPC systems in accelerating
scientific discovery by enabling complex and modern software stacks.


\begin{acks}
  This research is part of the Blue Waters sustained-petascale computing
  project, which is supported by the \grantsponsor{NSF}{National Science
  Foundation}{https://nsf.gov/} (awards \grantnum{NSF}{OCI-0725070} and
  \grantnum{NSF}{ACI-1238993}) and the State of Illinois. Blue Waters is a
  joint effort of the University of Illinois at Urbana-Champaign and its
  National Center for Supercomputing Applications.
  
  We thank CERN for the very successful operation of the LHC, as well as the
  support staff from \atlas institutions without whom \atlas could not be
  operated efficiently. The crucial computing support from all WLCG partners is
  acknowledged gratefully.  Major contributors of computing resources are
  listed in Ref~\cite{atlasthanks}.
  
  The authors gladly acknowledge valuable discussions with Edgar Fajardo,
  Stuart Anderson, and Peter Couvares. 
\end{acks}

\bibliographystyle{ACM-Reference-Format}
\bibliography{references}

\end{document}